\documentstyle[twoside,fleqn,espcrc2,epsf]{article}

\newcommand{\AmS}{{\protect\the\textfont2
  A\kern-.1667em\lower.5ex\hbox{M}\kern-.125emS}}

\title{Heavy quark mass dependence of semileptonic form factors for B
  decays\thanks{presented by S. Tominaga}} 

\author{JLQCD Collaboration\\
  S. Aoki\address{Institute of Physics, University of
    Tsukuba, Tsukuba, Ibaraki 305, Japan},
  M. Fukugita\address{Institute for Cosmic Ray Research, 
    University of Tokyo, Tanashi, Tokyo 188, Japan},
  S. Hashimoto\address{Computing Research Center,
    High Energy Accelerator Research Organization (KEK),\\
    Tsukuba, Ibaraki 305, Japan},
  K-I. Ishikawa\address{Department of Physics, Hiroshima
    University, Higashi-Hiroshima 739, Japan},
  N. Ishizuka$^{\rm a,e}$,
  Y. Iwasaki$^{\rm a,e}$,
  K. Kanaya$^{\rm a,}$\address{Center for Computational
    Physics, University of Tsukuba, Tsukuba, Ibaraki 305,
    Japan},\\ 
  Y. Kuramashi\address{Institute of Particle and Nuclear
    Studies, High Energy Accelerator Research Organization
    (KEK), Tsukuba, Ibaraki 305, Japan},
  H. Matsufuru$^{\rm d}$,
  M. Okawa$^{\rm f}$,
  T. Onogi$^{\rm d}$,
  S. Tominaga$^{\rm f}$,
  A. Ukawa$^{\rm a}$,
  N. Yamada$^{\rm d}$,\\
  T. Yoshi\'e$^{\rm a,e}$
}
       
\begin{document}

\begin{abstract}
  We present our study of the dependence of the
  heavy-to-light semileptonic $B$ decay form factors on the heavy-light 
  meson mass $M_{PS}$.  
  Simulations are made over a range of the heavy quark mass covering both 
  the charm and bottom quarks using the $O(a)$-improved clover action at 
  $\beta=5.9$ on a $16^3\times 40$ and $24^3\times 64$ lattice. 
  We find that a weak dependence of form factors on $M_{PS}$ observed in 
  previous studies in the region of charm quark persists up to the 
  region of $b$ quark.  
  The soft pion relation $f^0(q^2_{max})=f_B/f_\pi$ is examined and
  found to be largely violated.
\end{abstract}

\maketitle

\section{Introduction}

In spite of their small branching fraction, the exclusive
semileptonic decays $B\rightarrow\pi l\nu$ and
$B\rightarrow\rho l\nu$ are expected to become important 
processes to determine $V_{ub}$.  
A model independent calculation of the 
form factors relevant for these decay processes is possible 
using lattice QCD, and several attempts have already been 
made\cite{Reviews,BKS,APE,UKQCD}.
To avoid problems associated with large heavy quark mass $m_Q$, 
however, these studies made simulations in the charm 
quark mass region and extrapolated results to the $b$ quark. 

In this report we describe our examination of 
the heavy quark mass dependence of the form 
factors.  For this purpose we employ the $O(a)$-improved 
clover action in the formalism of the FNAL 
group\cite{El-Khadra_Kronfeld_Mackenzie_97}, and carry out 
simulations over a wide range of heavy quark mass including 
the bottom as well as charm quark masses\cite{Simone}. 
Our results for $f_B$ with this formalism show 
that systematic errors due heavy quark are small enough for 
the $b$ quark\cite{JLQCD_heavy-light_97}, and we
expect the systematic error to be also well under control
for the form factors.
 
\section{Simulation}

We use the plaquette action and the clover action 
with the clover coefficient determined at the one-loop 
level\cite{JLQCD_heavy-light_97}.  Measurements are made at 
$\beta=5.9$ with 100 configurations on a $16^3\times 40$ lattice 
and 110 configurations on a $24^3\times64$ lattice.
The lattice scale is set by the $\rho$ meson mass which gives
$a^{-1}=1.64(2)$~GeV.  Six values of the heavy quark hopping parameter 
in the range $\kappa_h=0.0718-0.1245$ are employed. 
The chiral limit is taken for the light quark 
with results obtained for the light quark hopping parameter 
in the range $\kappa_l=0.13630-0.13816$ to the critical value
$\kappa_c=0.13901(1)$.

\begin{figure}[tb]
  \vspace*{-4mm}
  \epsfxsize=7.5cm
  \epsfbox{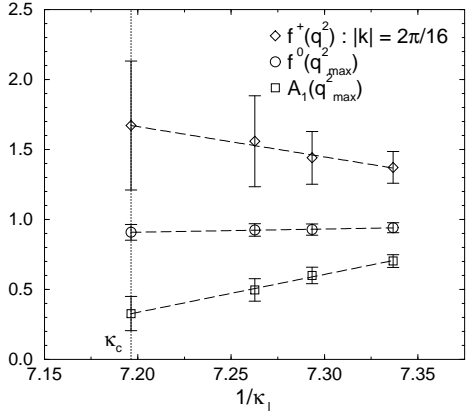}
  \vspace*{-12mm}
  \caption{Chiral extrapolation of $f^+(q^2)$ for minimum pion momentum, 
    $f^0(q^2_{max})$
    and $A_1(q^2_{max})$ obtained near $B$ mass region ($\kappa_h=0.0973$)
    for $16^3 \times 40$ lattice.}
  \label{fig:chiral_extrapolation}
\vspace*{-7mm}
\end{figure}

The form factors for the $B$ to $\pi$ decay are obtained
from the heavy-to-light three-point function 
$\langle P_{HL}(T/2)V_{\mu}(t)P_{LL}^{\dagger}(0)\rangle$ 
for the vector current $V_{\mu}(t)$ by dividing out 
the appropriate two-point functions.
The pseudo scalar fields $P_{HL}(T/2)$ and $P_{LL}(0)$ are 
smeared with the measured wave function obtained in 
Ref.~\cite{JLQCD_heavy-light_97}.
The $B$ meson is taken at rest
($|\vec{p}|$=0) and the pion is given up to a unit of 
momentum ($|\vec{k}|$=0, $2\pi/L$), above which the signal
becomes unacceptably noisy.
The form factors for the $B$ to $\rho$ decay is obtained 
in a similar manner, for which we consider only the
zero-recoil decay $|\vec{p}|=|\vec{k}|=0$.

We do not include the perturbative $Z$ factor for the 
currents in this study,
since results for finite heavy quark masses are not fully available.

\section{Results}

\begin{figure}[tb]
  \vspace*{-4mm}
  \epsfxsize=7.5cm
  \epsfbox{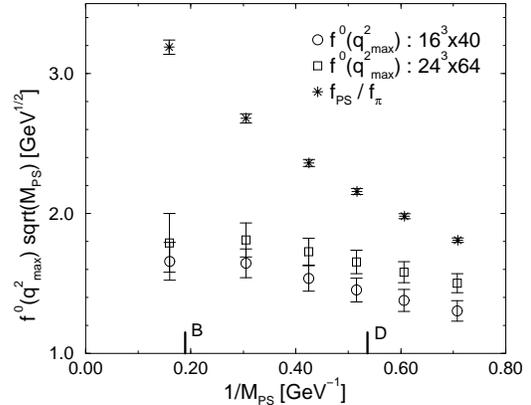}
  \vspace*{-12mm}
  \caption{$f^0(q^2_{max})\protect\sqrt{M_{PS}}$ as a
    function of $1/M_{PS}$. 
    Circles correspond to data on a $16^3\times40$
    lattice and squares on a $24^3\times64$ lattice.
    Crosses represent $f_{PS}\protect\sqrt{M_{PS}}/f_{\pi}$.
    }
  \label{fig:f0sqrtM}
\vspace*{-7mm}
\end{figure}

In Fig.~\ref{fig:chiral_extrapolation} we plot 
$f^+(q^2)$, $f^0(q^2_{max})$ and $A_1(q^2_{max})$ as a function of 
$1/\kappa_l$ for $\kappa_h=0.0973$ which is near the $b$ quark mass.  
Results for $f^+$ are for the minimal non-zero pion momentum.
We observe that the dependence in $1/\kappa_l$ is quite small for
$f^0(q^2_{max})$.  In contrast there is a clear slope for $A_1(q^2_{max})$, 
which differs from previous results\cite{BKS,APE}.
For the chiral extrapolation we adopt a linear form in $1/\kappa_\ell$ 
as shown in Fig.~\ref{fig:chiral_extrapolation}, 
with which our results are consistent.

Heavy quark symmetry predicts that 
$f^0\sqrt{M_{PS}}$, $f^+/\sqrt{M_{PS}}$ and
$A_1\sqrt{M_{PS}}$ scales toward the heavy quark
mass limit. 
In Fig.~\ref{fig:f0sqrtM} we plot
$f^0(q^2_{max})\protect\sqrt{M_{PS}}$ in the chiral limit
as a function of $1/M_{PS}$.
We observe that a small $1/M_{PS}$ correction, as suggested by results 
of previous studies performed around 
the charm quark mass\cite{Reviews,APE,UKQCD}, 
also holds in our data.  Furthermore the weak dependence persists to the 
region of $b$ quark mass. 
We also note a volume effect of about 10\% between $16^3\times40$ and 
$24^3\times64$ lattices, which is statistically significant toward 
light meson masses.

In the chiral limit, an application of the soft pion 
technique predicts the relation 
$f^0(q^2_{max})=f_{PS}/f_{\pi}$\cite{Soft_Pion_Theorem}. 
In Fig.~\ref{fig:f0sqrtM} we also plot $f_P\sqrt{M_{PS}}/f_{\pi}$ 
(crosses) obtained in Ref.~\cite{JLQCD_heavy-light_97}. 
There is a significant discrepancy with $f^0(q^2_{max})\sqrt{M_{PS}}$, 
particularly toward heavy quark masses.
Possible origins of the discrepancy are (i) subtleties in the chiral 
extrapolation of $f^0(q^2_{max})$ since $q^2_{max}$ changes with 
$1/\kappa_\ell$, 
(ii) systematic errors due to heavy quark including corrections from the 
$Z$ factor, (iii) scaling violation  
and breaking of chiral symmetry for light quark.  It is not clear at present
if these could account for the large difference seen in Fig.~\ref{fig:f0sqrtM}.

\begin{figure}[tb]
  \vspace*{-4mm}
  \epsfxsize=7.5cm
  \epsfbox{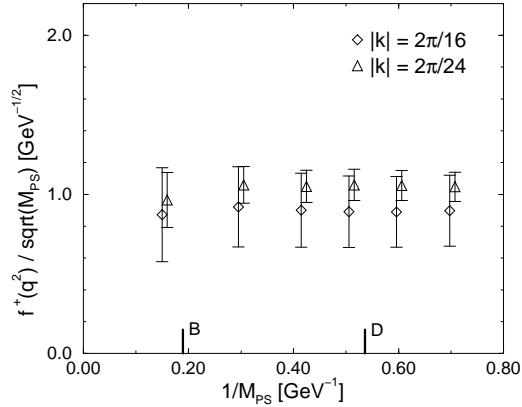}
  \vspace*{-9mm}
  \caption{$f^+(q^2)/\protect\sqrt{M_{PS}}$ in the chiral
    limit as a function of $1/M_{PS}$. 
    Triangles correspond to the results on a $16^3 \times
    40$ lattice and diamonds on a $24^3 \times 64$ lattice. 
    $B$ meson is at rest and pion carries one unit of momentum.
    }
  \label{fig:fp}
\vspace*{-7mm}
\end{figure}

The $1/M_{PS}$ dependence of $f^+(q^2)/\sqrt{M_{PS}}$ is shown in
Fig.~\ref{fig:fp}.  It is negligible for this quantity.

\begin{figure}[tb]
  \vspace*{-4mm}
  \epsfxsize=7.5cm
  \epsfbox{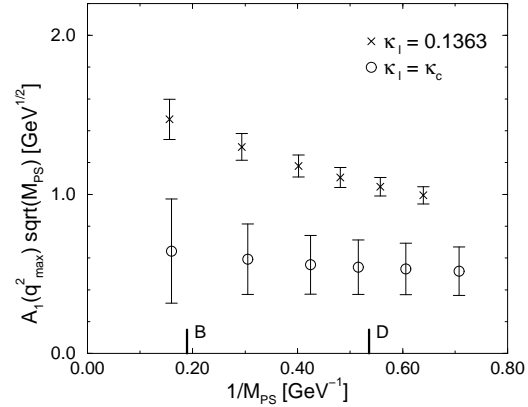}
  \vspace*{-9mm}
  \caption{$A_1(q^2_{max})\protect\sqrt{M_{PS}}$ on a
    $16^3\times40$ lattice at $\kappa_l=0.1363$ and $\kappa_c$ as a
    function of $1/M_{PS}$. }
  \label{fig:A1}
\vspace*{-7mm}
\end{figure}

For $f^0$ and $f^+$ the slope in the chiral extrapolation is
almost independent of the heavy quark mass. 
Therefore the $1/M_{PS}$ dependence of these form factors 
is almost unchanged by the chiral extrapolation,
as was seen in Ref.~\cite{APE}.
On the other hand, the large slope for $A_1$ shown in 
Fig.~\ref{fig:chiral_extrapolation} becomes smaller as the heavy 
quark becomes lighter.
As a result the $1/M_{PS}$ behavior of $A_1$ depends
significantly on the light quark mass as shown in 
Fig.~\ref{fig:A1}, where we plot results at a finite light
quark mass (crosses) together with that in the chiral limit(circles).

\section{Conclusions}

Previous studies\cite{Reviews,APE,UKQCD} performed on 
the charm quark mass region suggested that the $1/M_{PS}$
dependence of the form factors is small contrary to
the case of the heavy-light decay constant which
varies significantly between the static limit and the region of charm 
quark.
Our calculation confirms this trend and extends it to the region of 
$b$ quark.  
This is particularly puzzling for $f^0$ for which we find 
a large violation of the soft pion relation 
$f^0(q^2_{max})=f_B/f_{\pi}$.  A weak $1/M_{PS}$ dependence 
and a discrepancy from the soft pion relation are 
also observed if non-relativistic QCD is employed for heavy 
quark\cite{hiroshima}. We feel that further understanding 
of the heavy quark mass dependence is required for a reliable calculation 
of heavy-to-light form factors from lattice QCD.
\vspace{2mm}\\
This work is supported by the Supercomputer 
Project (No.~97-15) of High Energy Accelerator Research Organization (KEK),
and also in part by the Grants-in-Aid of 
the Ministry of Education (Nos. 08640349, 08640350, 08640404,
09246206, 09304029, 09740226).
Two of us (H.M. and S.T.) are supported 
by the JSPS Research Fellowships.

\end{document}